\documentclass[aps,prb,superscriptaddress,amsmath,amssymb,floatfix,twocolumn]{revtex4-1}

\usepackage{times}
\usepackage{bm}
\usepackage{graphicx,graphics,color}
\usepackage[colorlinks,citecolor=blue,linkcolor=red]{hyperref}
\usepackage{color}

\begin{document}

\title{Surface Local Impurity Scattering as a Probe for Topological Kondo Insulators}

\author{C.-C. Joseph Wang}
\affiliation{Theoretical Division and Center for Nonlinear Studies, Los Alamos National Laboratory,
Los Alamos, New Mexico 87545, USA}
\altaffiliation{Current address: 
Quantum Computational Science Group, Quantum Information Science Section, Oak Ridge National Laboratory, 1 Bethel Valley Road, Oak Ridge, Tennessee 37831, USA}

\author{Jean-Pierre Julien}
\affiliation{Institut N\'{e}el CNRS \& Universit\'{e} de Grenoble Alpes, 25 Avenue des Martyrs, BP 166, F-38042
Grenoble Cedex 9, France }

\author{A. V. Balatsky}
\affiliation{Institute for Materials Science, Los Alamos National Laboratory,
Los Alamos, New Mexico 87545, USA}

\author{Jian-Xin Zhu}
\email[
Electronic address: ]{jxzhu@lanl.gov}
\affiliation{Theoretical Division and Center for Nonlinear Studies, Los Alamos National Laboratory,
Los Alamos, New Mexico 87545, USA}
\affiliation{Center for Integrated Nanotechnologies, Los Alamos National Laboratory, Los Alamos, New Mexico 87545, USA}

\begin{abstract}
Shortly after the discovery of topological band insulators, topological Kondo insulators (TKIs) have also been theoretically predicted. The latter has ignited revival interest in the properties of Kondo insulators. Currently, the feasibility of topological  nature in SmB$_6$ has been intensively analyzed by several complementary probes.  Here by starting with a minimal-orbital Anderson lattice model, we explore the local electronic structure in a Kondo insulator. We show that the two strong topological regimes sandwiching the weak topological regime give rise to a single Dirac cone, which is located near the center or corner of the surface Brillouin zone. We further find that, when a single impurity is placed on the surface, low-energy resonance states are induced in the weak scattering limit for the strong TKI regimes and the resonance level moves monotonically across the hybridization gap with the strength of impurity scattering potential; while low energy states can only be induced in the unitary scattering limit for the weak TKI regime, where  the resonance level moves universally toward the center of the hybridization gap. These impurity-induced low-energy quasiparticles will lead to characteristic signatures in scanning tunneling microscopy/spectroscopy, which has recently found success in probing into exotic properties in heavy fermion systems. 
\end{abstract}

\maketitle

\section{Introduction} 
Topological insulators (TIs) are a novel state of matter~\cite{MZHasan:2010,XLQi:2010,XLQi:2011}. Different from conventional band insulators,  the new class of materials exhibit not only a bulk insulating band gap  but also gapless spin-filtered edge states in two-dimensions or a metallic Dirac fermion surface states in three dimensions. So far, the compounds HgTe, Bi$_2$Se$_3$, Bi$_{1-x}$Sb$_x$, Bi$_2$Te$_3$, and TlBiTe$_2$ have been identified as weakly interacting TIs~\cite{YXia:2009,PRoushan:2009,TZhang:2009,JSeo:2010,ZAlpichshev:2010}, where the band inversion is driven by  the spin-orbit coupling~\cite{BABernevig:2006}. The main complication in these band insulators (especially Bi-based compounds) is that they still have a high concentration of bulk carriers, giving rise to a considerable residual conductivity in the sample bulk. Soon after their discovery, the possibility of interaction driven topological insulators~\cite{SRaghu:2008,KSun:2009,RNandkishore:2010,KSun:2012,HMGuo:2009,DAPesin:2010,XWan:2011,BJYang:2010,MDzero:2010,XZhang:2012,BYan:2012,MDzero:2012,MTTran:2012,FLu:2013,XDeng:2013,VAlexandrov:2013,XYFeng:2013} has been discussed in the context that the bulk insulating  behavior may be enhanced by the interplay between electron correlation and spin-orbit coupling. 
In particular, there has been intensified interest in the possibility of topological Kondo insulators in $f$-electron materials. Kondo insulators are a type of heavy fermion materials that have been studied for nearly four decades. In these materials, the on-site Coulomb repulsion on localized $f$-electrons significantly renormalizes the $f$-electron band width. Theoretically, the idea of TKIs has been first proposed by Dzero and co-workers, who showed that Kondo insulators could develop topologically nontrival ground states, which are connected adiabatically to those in noninteracting insulators~\cite{MDzero:2010}.  The proposal has sparked a flurry of experimental attempts to demonstrate the existence of topological surface states on a prototype of Kondo insulators, SmB$_6$, which is a stoichiometric compound  (distinct from the topological band insulators as mentioned above) and has recently been suggested to be a class of topological Kondo insulators (TKIs)~\cite{TTakimoto:2011}. On the one hand, the notion of TKI has been adopted to explain results from several transport~\cite{SWolgast:2013,ZJYue:2013}, de Haas-van Alphen effect~\cite{GLi:2013}, angle-resolved photoemission spectroscopy~\cite{NXu:2013,MNeupane:2013,JJiang:2013}, and scanning tunneling microscopy (STM)~\cite{MMYee:2013}  experiments. On the other hand,  the ARPES has not unanimously agreed upon the associated Dirac cones while it has also been proposed that the resistivity in SmB$_6$ can arise through native surface instabilities such as a non-TI metallic surface states~\cite{ZHZhu:2013}, a band inversion layer~\cite{EFrantzeskakis:2013}, or surface reconstruction~\cite{SRobler:2013}.
Therefore, more theoretical and experimental efforts are needed to uncover the features manifesting the topological aspect of SmB$_6$ in particular and other interesting Kondo insulators in general.  The use of single impurity with the STM technique has proved to be a powerful approach to distinguish the nature of underlying electronic states in strongly correlated electron systems. It has been used to  identify the pairing symmetry in unconventional 
superconductors~\cite{AVBalatsky:2006,JXZhu:2011,CLSong:2013}, and to shed insight into the hidden order state in URu$_2$Si$_2$~\cite{MHHamidian:2011}.

In  this work,  we study the local electronic structure around a single impurity on the surface of a topological Kondo insulator. Within the Gutzwiller method, we are able to establish the necessary condition for the existence of a bulk energy gap {\em near the Fermi energy}. On top of these bulk states, introducing a single impurity scatter on the surface of Kondo insulators generates local electronic response dramatically different from that in trivial semiconductors. We further find that the existence of impurity induced bound state is sensitive to the potential strength. This dependence is unique to the topological nature of the Kondo insulator state, out of which the surface metallic state emerge.  The prediction should be readily accessible to the STM experiments, in view of the recent success of this technique applied to understand the bulk and surface properties of SmB$_6$~\cite{MMYee:2013,SRobler:2013}.

The outline of this work is as follows: In Sec.~\ref{sec:results}, the dependence of the topological properties on the model parameter is studied. The relation between the number of Dirac cone in the surface Brillouin zone and the topological indices is discussed. Furthermore, the existence of the low-energy quasiparticle states around a single impurity is elucidated in both strong and weak topological insulator regimes. A summary is given in Sec.~\ref{sec:summary}.   In Sec.~\ref{sec:method}, we describe  a generalized Anderson lattice model, the Gutzwiller variational wavefunction approach, and numerical details. The characterization of $Z_2$ topological indices and the $T$-matrix method for the calculation of local electronic structure around a nonmagnetic impurity are presented in the same section.

\section{Surface Electronic Structure and Quasiparticle States around a Single Impurity}
\label{sec:results}

\begin{figure}[h!]
\centering\includegraphics[
width=1.0\linewidth,clip]{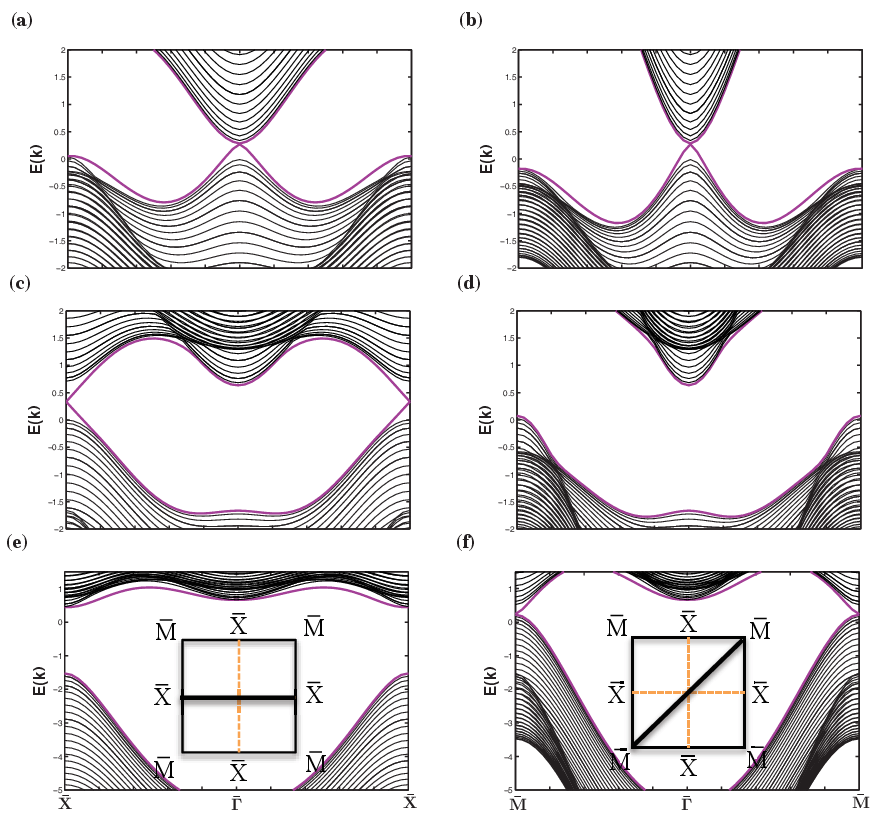}
\caption{The band dispersion in STKI-I (a)-(b), WTKI (c)-(d), and STKI-II (e)-(f) states along the bond (10)  (left panels) and the diagonal (11) (right panels) directions of surface Brillouin zone.  The purple curves mark the surface states.
}
\label{FIG:dispersion}
\end{figure}

\subsection{Surface band structure in the pristine case.}
\begin{figure}[t!]
\centering\includegraphics[
width=1.0\linewidth,clip]{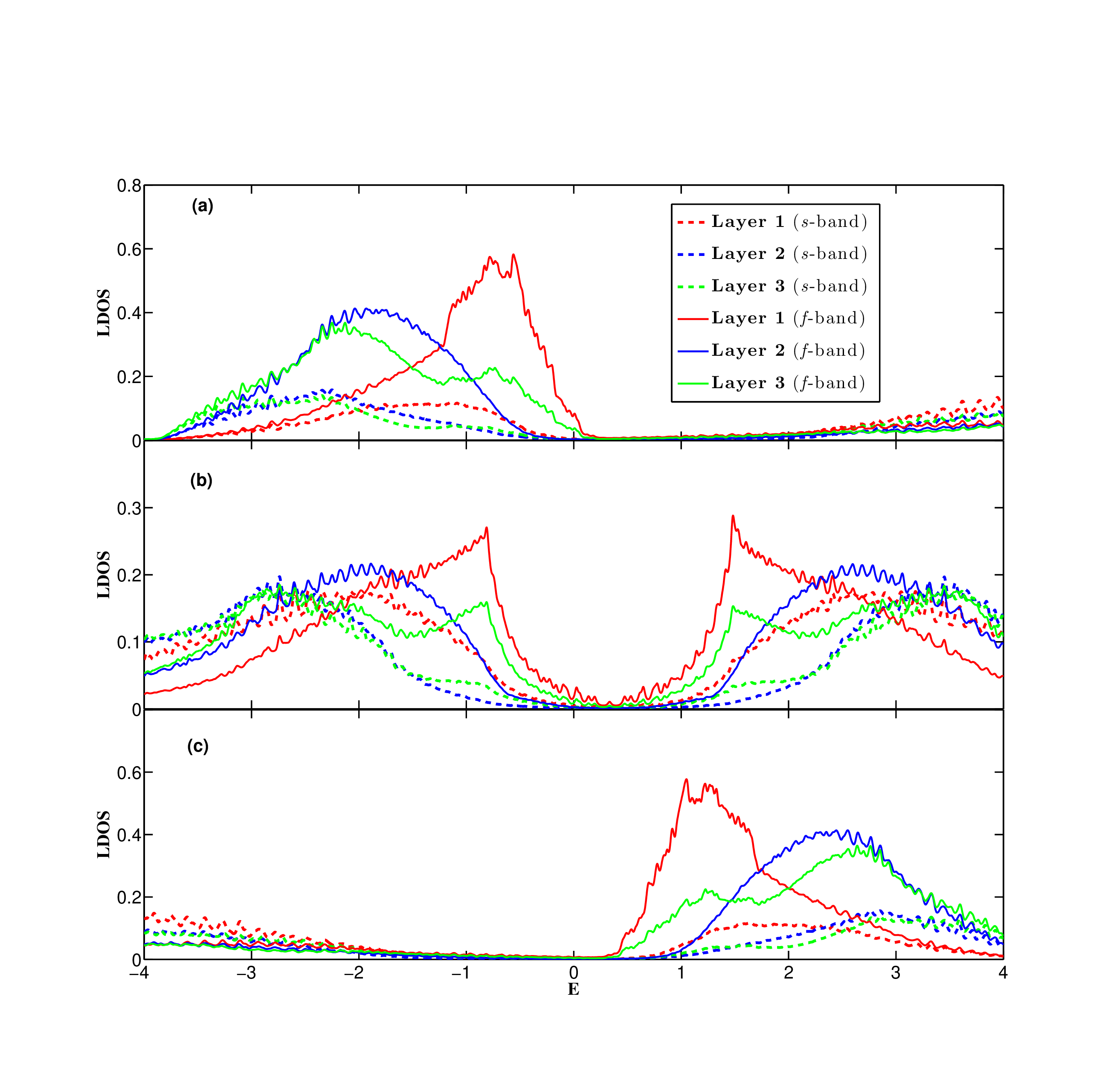}
\caption{Local density of states near the surface of the Kondo insulator in the STKI-I (a), WTKI (b), and STKI-II (c) regimes. Within each panel, the projected-$s$ and $f$ local density density of states per spin in the first three top layers in a slab structure are shown.  
}
\label{FIG:surf_ldos}
\end{figure}

We begin with the electronic structure on the surface of the Kondo insulator.  To do so, we have determined the topological characteristics of the Kondo insulating state in the  system bulk through the calculation of the $Z_2$  topological indices.  Dependent on the location of the localized $f$-level $\epsilon_f$, the Kondo insulating states can be characterized into two strong topological insulating (STKI) regimes, which are separated by  one weak topological insulating (WTKI) regime~\cite{MDzero:2010}.  Hereafter we call the two STKI regimes STKI-I and STKI-II. We note the topological structure has been systematically studied in previous model calculations~\cite{MTTran:2012,MDzero:2012}, and we refer to these works for bulk band structure.
 Here we pay more attention to the existence of a bulk energy gap near the Fermi energy, which is a necessary condition for a truly insulating bulk. Distinctly, we find that this condition can be satisfied by adjusting the chemical potential self-consistently so that  the whole system is close to half filled in our minimal two-orbital model. Thereafter, we present results with the values of bare $f$-level to be $\epsilon_{f} = -17$, $-7.5$, and $2$ to represent the STKI-I, WTKI, and STKI-II regimes.  In Fig.~\ref{FIG:dispersion}, the energy dispersion is shown along the in-plane bond and diagonal directions of  the surface Brillouin zone  (BZ) for the pristine slab geometry. It can be seen that for the WTKI state, there are two Dirac cones located at the $\bar{X}$ points, that is, the edges of the surface BZ. For the STKI-I state, there exists a Dirac cone at the $\bar{\Gamma}$ point, that is, the center of the surface BZ; while for the STKI-II state, there exists a Dirac cone at the $\bar{M}$ points, that is, the corner of the surface BZ.  To gain further insight into the structure of the surface metallic states, we show in Fig.~\ref{FIG:surf_ldos} the local density of states in the first three surface planes of the slab geometry. Noticeably, the insulating gap for all three regimes is open near the Fermi energy, which makes the notion of Kondo insulator really meaningful.  In the STKI-I state, we can see that most of the $f$-electron spectral weight is lumped at the lower edge of the insulating gap, implying a dominant electrons occupation of the $f$-band; while in the STKI-II state, we see that most of the $f$-electron spectral weight is lumped at the upper edge of the insulating gap, implying a minor electron occupation of the $f$-band.  In the WTKI state, the $f$-electron spectral weight is about equally distributed at the lower and upper edges of the energy gap. 
These unique distribution of the $f$-electron spectral weight near the gap edges seems to be related to how the particle-hole symmetry is breaking, which controls
 the number of Dirac cones and their locations in the three TKI regimes.  Furthermore, one can also see that the dominant peaks in the local density of states at the surface are located closer to the Fermi energy than those deep into the bulk. This band gap narrowing indicates that the Kondo coherence is weakened at the surface as compared to that in the bulk, arising from the disruption of the nearest-neighbor $c$-$f$ hybridization along the direction perpendicular to the surface. The result is also consistent with a recent proposal of surface Kondo breakdown in topological Kondo insulators~\cite{OErten:2016}.

\begin{figure}[t!]
\centering
\includegraphics[width=1.0\linewidth,clip]{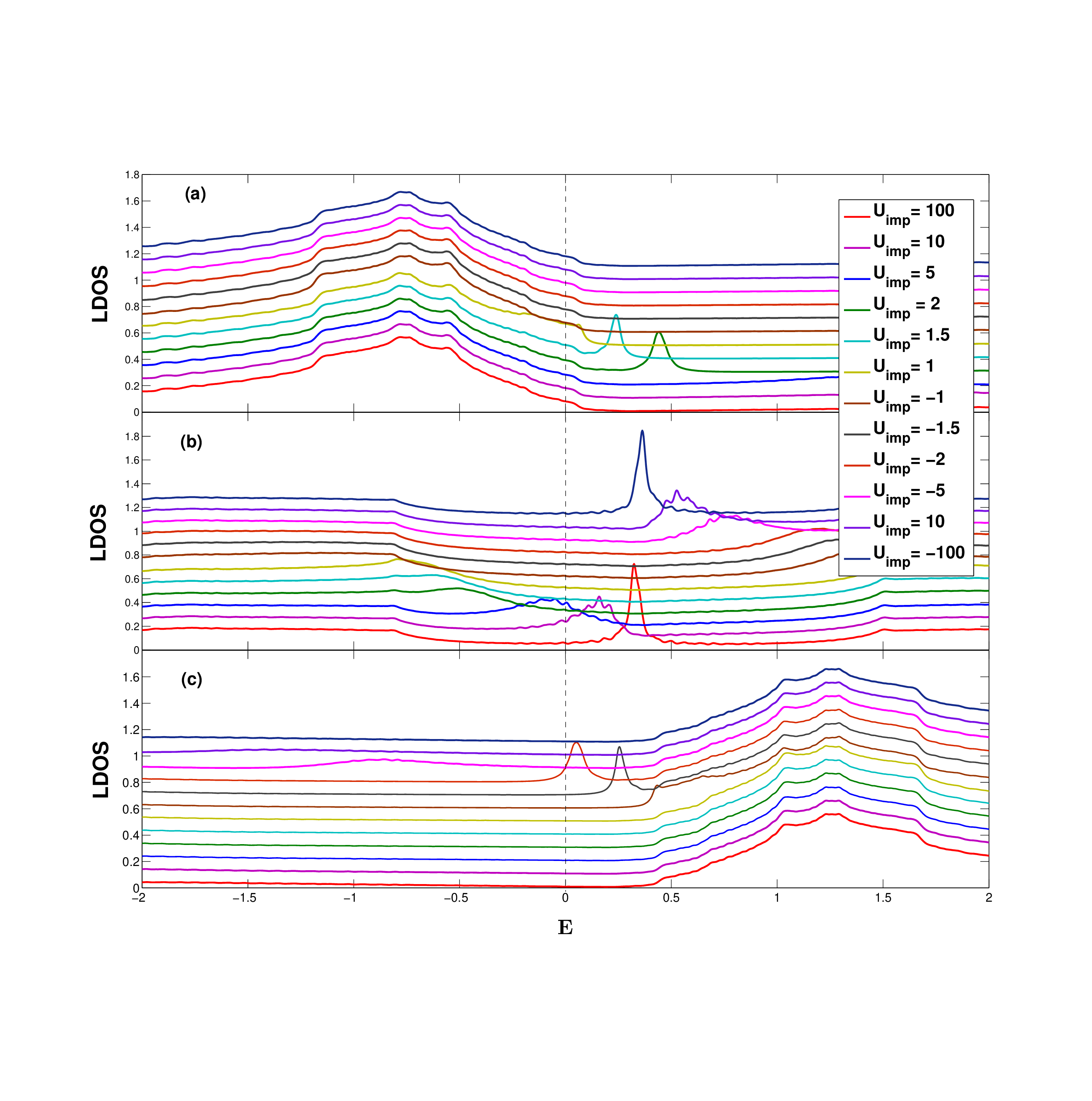}
\caption{The LDOS for the $f$-electron per spin at a site nearest-neighboring to the single impurity imbedded in on the top layer of the Kondo insulator slab. The bulk states are in  the  STKI-I (a), WTKI (b), and STKI-II (c) regimes. Within each panel, the LDOS for various values of impurity potential strength is plotted while an offset of 0.1 between two consecutive curves of LDOS is chosen for clarity. 
}
\label{FIG:imp_dos}
\end{figure}

\begin{figure}[h!]
\centering
\includegraphics[width=1.0\linewidth,clip]{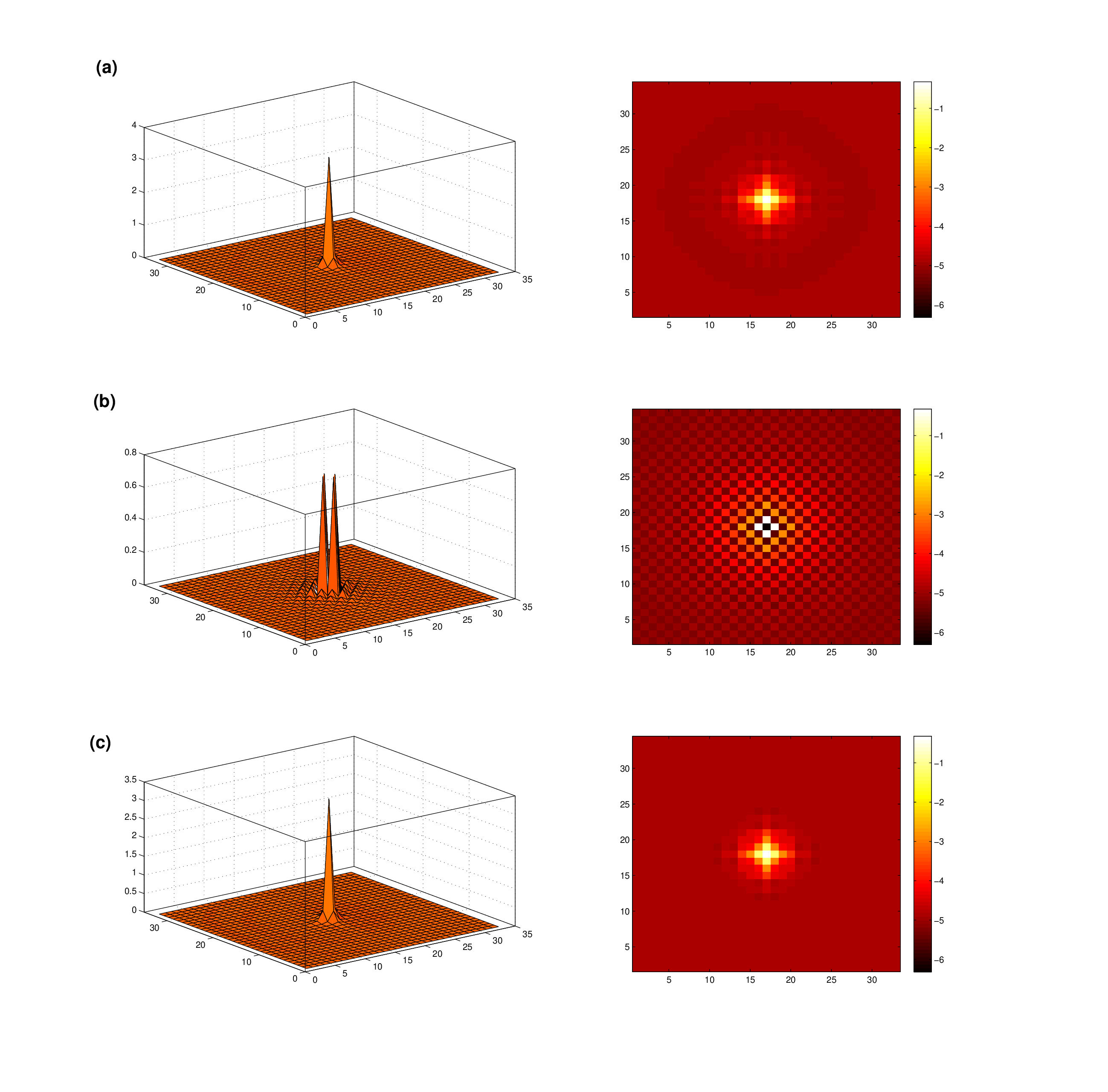}
\caption{The spatial dependence of the LDOS  for the $f$-electron per spin at the resonance energy around a single impurity imbedded at the top layer of the  Kondo insulator slab. The left panels are for  the surface plots while the right ones are for the  image plots with the intensity in log scale. The intra-gap impurity resonance state energy is determined by the strength of the impurity potential in the respective STKI-I (a) [$E_r=0.441$ for $U_{\textrm{imp}}=2.0$], WTKI (b) [$E_r=0.324$ for $U_{\textrm{imp}}=100$], and STKI-II (c) [$E_r=0.051$ for $U_{\textrm{imp}}=-2.0$] regimes. 
}
\label{FIG:imp_image}
\end{figure}

\subsection{Local electronic structure around a single impurity at the surface}
We next explore the effect of a single impurity on the surface of the Kondo insulator. As we have mentioned in the introduction, a study of the local electronic properties around these single impurities will help elucidate the underlying feature of the electron states.  They are especially effective in a wide range of Dirac materials including $d$-wave superconductors, graphene, and surface metallic states in band topological insulators.  Figure~\ref{FIG:imp_dos} shows the local density of states (LDOS) around a single nonmagnetic impurity for three regimes of Kondo insulating state. The LDOS is measured at a site nearest neighboring to the impurity site. For each regime, the LDOS for the $f$-electron per spin is shown for a sequence of values of impurity potential strength from the repulsive unitary limit to the attractive unitary limit. For the STKI-I regime, the intragap impurity resonance peak occurs when the strength of the impurity potential is around $U_{\textrm{imp}}$=2.0, 1.5, and 1.0.  Noticeably, the intragap resonance peak is shifted monotonically from positive energy side to the negative energy side and then merges with the spectrum continuum at the lower edge of Kondo insulating gap (see Fig.~\ref{FIG:imp_dos}(a)), when the impurity potential is changed from the positive unitary limit toward the negative unitary limit. Correspondingly, for the STKI-II regime, the intragap impurity resonance peak occurs when the strength of the impurity potential is around $U_{\textrm{imp}}=-2.0$, $-1.5$, and $-1.0$. Now the intragap resonance peak is shifted monotonically from negative  energy side to the positive energy side and then merges with the spectrum continuum at  the upper edge of Kondo insulating gap (see Fig.~\ref{FIG:imp_dos}(c)), when the impurity potential is changed from the negative unitary limit toward the positive unitary limit. Therefore, an interesting dual relation between the STKI-I and STKI-II regimes is established. We emphasize here that in the strong topological Kondo insulating regimes, there exist no impurity resonance states in the unitary limit of impurity scattering. On the contrary, in the WTKI regime, a re-entrant behavior of the impurity resonance state is obtained. The impurity resonance peak is pinned on the lower edge side of the  Kondo insulating gap center (see the bottom red line of Fig.~\ref{FIG:imp_dos}(b)) in the unitary limit of a repulsive impurity potential. With the decreasing repulsive impurity potential, the peak is shifted toward the negative energy side of the $f$-electron spectrum continuum.  After the impurity potential changes sign from being repulsive into being attractive, the resonance peak emerges out of the $f$-electron spectrum continuum  with the increased strength of the potential scattering. This peak is finally pinned on the upper edge side of the gap center as the unitary limit of an attractive impurity potential is reached (see the top black line of Fig.~\ref{FIG:imp_dos}(b)). Therefore, for the WTKI regime, the most salient feature is that there exist impurity resonance states in the unitary limit of impurity scattering regardless of being repulsive or attractive. 

The different response of electronic states to the impurity scattering on the surface of different Kondo insulating regimes can also be revealed in the spatial dependence of $f$-electron LDOS around the impurity. As shown in Fig.~\ref{FIG:imp_image}, for the impurity induced resonance states sustained only at the weak scattering of impurity potential in the STKI regimes, the LDOS intensity is spatially peaked at the impurity site itself; while the resonance states  sustained at the unitary limit of impurity scattering for the WTKI regime, the LDOS intensity is depressed significantly at the impurity site and exhibits maxima instead at the four sites nearest neighboring (along the bond direction) to the impurity site. 

The very different outcome of impurity scattering  in the STKI and WTKI regimes, especially the absence  of impurity resonance states in the STKI regimes versus the presence of the impurity resonance states in the WTKI regime in the unitary limit of impurity scattering,  is believed to be related to the number of the Dirac cones hosted by the surface metallic states, which support only one Dirac cone in the STKI regimes while two Dirac cones in the WTKI regime. On the one hand, our observation for the WTKI regime, with the two Dirac cones on the surface,  of the existence of the impurity resonance states in the unitary impurity scattering limit, agrees well with the generic feature of impurity scattering in other types of Dirac materials, where an even number of Dirac cones exist. In particular, we notice that the impurity resonance states have been obtained in the unitary limit of impurity scattering in both two-dimensional $d$-wave superconductors with four Dirac cones  and  graphene with two Dirac nodes. On the other hand, our observation in the STKI regimes of no impurity resonance states in the unitary impurity scattering limit echoes the expectation that surface metallic states  with an odd number of gapless chiral modes supported by a strong topological insulator are enjoying topological protection against surface impurity scattering. Thanks to the stoichiometric nature of the heavy fermion compounds, we expect that the emerging topological Kondo insulators can provide an unprecedented test ground for the rich results uncovered for the surface impurity scattering effect. Although there is no STM study so far to focus on these novel effects from the impurity scattering for the purpose of identifying the topological nature of Kondo insulators, there is equally no reason to exclude such a powerful approach. 

\section{Concluding remarks}
\label{sec:summary}
In summary, we have studied  the local electronic structure in a Kondo insulator within a minimal-orbital Anderson lattice model.
We have shown first that the two strong topological regimes sandwiching the weak topological regime give rise to a single Dirac cone, which is located near the center or corner of the surface Brillouin zone. We have further found that, when a single impurity is placed on the surface, low-energy resonance states are induced in the weak scattering limit for the strong TKI regimes and the resonance level moves monotonically across the hybridization gap with the strength of impurity scattering potential; while low energy states can only be induced in the unitary scattering limit for the weak TKI regime, where  the resonance level moves universally toward the center of the hybridization gap. These results on the impurity induced low-energy quasiparticles should be detectable in scanning tunneling microscopy/spectroscopy, which has recently found success in probing into exotic properties in heavy fermion systems.


\section{Theoretical Methods}
\label{sec:method}

Our starting point is a generalized periodic Anderson model
\begin{equation} 
H=H_{0} + H_{\textrm{imp}}\;.
\label{eq:hamil}
\end{equation}
The term on the right-hand side (RHS) of Eq.~(\ref{eq:hamil}), $H_0$, describes the pristine bulk or slab structure of the heavy fermion system and consists of three parts: 
\begin{equation}
H_{c} = - \sum_{ij,\sigma} (t_{ij}^{c} + \mu \delta_{ij})c_{i\sigma}^{\dagger} c_{j\sigma}\;,
\end{equation}
for the conduction electrons, 
\begin{equation}
H_{f} = \sum_{i,\alpha} (\epsilon_{f}-\mu) f_{i\alpha}^{\dagger} f_{i\alpha}
+ \sum_{i} U_{f} n_{i\uparrow}^{f} n_{i\downarrow}^{f}+ (J_H/2)\sum_{ij} \mathbf{S}_i \cdot \mathbf{S}_j\;,
\end{equation}
 for local $f$-electrons,  and 
\begin{equation}
H_{hyb}=\sum_{ij,\sigma\alpha}  [ c_{i\sigma}^{\dagger}V_{cf,ij,\sigma\alpha} f_{j\alpha} + \textrm{H.c.}]\;,
\end{equation}
for hybridization between the conduction and $f$-electrons. 
The second term on the RHS of Eq.~(\ref{eq:hamil}),
\begin{equation} 
H_{\textrm{imp}}= \sum_{\alpha} U_{imp} f_{0\alpha}^{\dagger} f_{0\alpha} \;,
\end{equation}
describes the 
single-site impurity scattering, which without loss of generality is located at the origin of the lattice coordinate system.
Here the operators $c_{i\sigma}^{\dagger}$ ($c_{i\sigma}$) create (annihilate) a conduction electron at site $\mathbf{r}_{i}$ with spin projection $\sigma$ while the operators $f_{i\alpha}^{\dagger}$ 
($f_{i\alpha}$) create (annihilate) a $f$-level electron at site $\mathbf{r}_{i}$ with pseudo-spin projection $\alpha$ representing the $\Gamma_{8(2),+}$ and $\Gamma_{8(2),-}$ components due to strong spin-orbit coupling~\cite{MTTran:2012}.
The number
operators for $c$ and $f$ orbitals with spin projection are given by
$n_{i\sigma}^{c}=c_{i\sigma}^{\dagger}c_{i\sigma}$ and
$n_{i\alpha}^{f}=f_{i\alpha}^{\dagger}f_{i\alpha}$,
respectively. 
The quantity $t_{ij}^{c}$ is the hopping integral of the conduction electrons,  $\epsilon_{f}$ is the local $f$-orbital energy level on the magnetic atoms, and $\mu$ is the chemical potential.
The hybridization matrix between the conduction band and $f$-orbital on the magnetic atoms is represented by $V_{cf,ij}$ and the $f$-electrons on the magnetic atoms 
experience the Coulomb repulsion of strength $U_f$.  For a simple cubic lattice, by considering the hybridization between 6$s$ and 4$f$  electrons of rare-earth systems in the limit of strong spin-orbit coupling for the $f$-electrons,  one can model  the nearest-neighbor three-dimensional hybridization in the form:~\cite{MTTran:2012}
\begin{equation}
V_{cf,ij} = \mathbf{d}_{ij}\cdot \bm{\sigma}\;.
\end{equation}
Here $\bm{\sigma}$ is the Pauli matrix and $\mathbf{d}_{ij}= (V_x d_{ij}^{x}, V_y d_{ij}^{y}, V_z d_{ij}^{z})$, where $V_x=V_y=V_z/2=V_{cf}$, while $d_{ij}^{x} = \delta_{i_y,j_y} \delta_{i_z,j_z} (x_i-x_j)/a$, $d_{ij}^{y} = -\delta_{i_x,j_x} \delta_{i_z,j_z} (y_i-y_j)/a$, and $d_{ij}^{z} = -\delta_{i_x,j_x} \delta_{i_y,i_y} (z_i-z_j)/a$. The quantity $a$ is the lattice constant of the simple cubic system. We note that although for the typical systems like SmB$_6$, the dominant hybridization could occur between 5$d$ and 4$f$ orbitals, the hybridization structure should be similar and the  essential physics as described in the present work should be robust. 
In addition, since the Gutzwiller method (see Sec. 4.1) is a local approximation, which does not capture the conduction-electron mediated Ruderman-Kittel-Kasuya-Yoshida (RKKY) exchange interaction, we have also introduced a Heisenberg-like exchange interaction to model the spin liquid~\cite{TSenthil:2003,TSenthil:2004,CPepin:2007,JXZhu:2008}. This type of effective models have been most studied for the non-Fermi-liquid physics near the heavy-fermion quantum criticality~\cite{PGegenwart:2008}.  We generalize the Gutzwiller approximation~\cite{JXZhu:2012} to solve the pristine part of the Hamiltonian $H_0$ and proceed to study the local electronic structure around the single impurity within the $T$-matrix method~\cite{AVBalatsky:2006}.

\subsection{Gutzwiller approximation method}
Due to the presence of onsite Hubbard interaction $U_{f}$ between the $f$-electrons on each lattice site in Eq.~(\ref{eq:hamil}), the problem even for the pristine system $H_0$ is already strongly correlated. This strong correlation
effect can be accounted for by reducing the statistical weight of double occupation in the Gutzwiller projected wavefunction approach~\cite{MCGutzwiller:1963}, and the projection can be carried out semi-analytically within the Gutzwiller approximation~\cite{MCGutzwiller:1965,DVollhardt:1984,FCZhang:1988}.  In the present problem, the lattice translation symmetry is already broken
in the slab geometry, we thus use a spatially unrestricted Gutzwiller approximation~\cite{CLi:2006,JPJulien:2006,QHWang:2006,WHKo:2007,NFukushima:2008} to translate the Hamiltonian $H_0$ for the pristine system into the following renormalized mean-field Hamiltonian:
\begin{widetext}
\begin{eqnarray}
H_{\textrm{eff},0}&=& - \sum_{ij,\sigma} (t_{ij}^{c} + \mu \delta_{ij})c_{i\sigma}^{\dagger} c_{j\sigma}  + \sum_{ij,\sigma\alpha}  [V_{cf,ij,\sigma\alpha} g_{t,j\alpha}  c_{i\sigma}^{\dagger} \tilde{f}_{i\alpha} + \textrm{H.c.}] 
 + \sum_{i,\alpha} (\epsilon_{f}+\lambda_{i} -\mu) \tilde{f}_{i\alpha}^{\dagger} \tilde{f}_{i\alpha} \nonumber \\
 &&-\frac{3J_{H}}{4}\sum_{ij,\alpha} g_{s,i}g_{s,j} \tilde{\chi}_{ij} \tilde{f}_{i\alpha}^{\dagger} \tilde{f}_{j\alpha}  + \sum_{i} U_{f} d_{i} \;,
\label{EQ:Hamil_GA}
\end{eqnarray}  
\end{widetext}
where $\lambda_{i}$ and $d_{i}$ are the Lagrange multiplier and the double occupation 
at site $i$.  We have used $\tilde{f}_{i\alpha}^{\dagger}$ ($\tilde{f}_{i\alpha}$) to denote the quasiparticle field operators to differentiate from the truly $f$-electron operators in Eq.~(\ref{eq:hamil}). The spin liquid term is described by the resonant valence bond order $\tilde{\chi}_{ij} = \sum_{\alpha} \tilde{f}_{i\alpha}^{\dagger} \tilde{f}_{j\alpha}/2$. The local  $c$-$f$ hybridization and spin-exchange interaction have been renormalized by a factor of $g_{t,i\alpha}$ and $g_{s,i}g_{s,j}$, respectively. These $g$ factors are given by 
\begin{equation}
g_{t,i\alpha} = \biggl{[} \frac{(\bar{n}_{i\alpha}^{\tilde{f}} -d_i)(1-  \bar{n}_{i}^{\tilde{f}}  + d_{i})}
{\bar{n}_{i\alpha}^{\tilde{f}}  (1-\bar{n}_{i\alpha}^{\tilde{f}} )}\biggr{]}^{1/2}
+ \biggl{[} \frac{d_i(\bar{n}_{i\bar{\alpha}}^{\tilde{f}}  - d_{i})}
{\bar{n}_{i\alpha}^{\tilde{f}}  (1-\bar{n}_{i\alpha}^{\tilde{f}} )}\biggr{]}^{1/2}\;,
\label{eq:gt}
\end{equation} 
and 
\begin{equation}
g_{s,i} = \biggl{[} \frac{(\bar{n}_{i\bar{\alpha}}^{\tilde{f}} -d_i)(\bar{n}_{i\alpha}^{\tilde{f}} -d_i)}
{\bar{n}_{i\bar{\alpha}}^{\tilde{f}}  (1-\bar{n}_{i\alpha}^{\tilde{f}})
\bar{n}_{i\alpha}^{\tilde{f}}  (1-\bar{n}_{i\bar{\alpha}}^{\tilde{f}})}\biggr{]}^{1/2}\;.
\label{eq:gs}
\end{equation} 
Here $\bar{n}_{i\alpha}^{\tilde{f}}$ being the expectation value of the pseudo spin-$\alpha$ density operator $n_{i\alpha}^{\tilde{f}}=\tilde{f}_{i\alpha}^{\dagger} \tilde{f}_{i\alpha}$
and $\bar{n}_{i}^{\tilde{f}}=\sum_{\alpha} \bar{n}_{i\alpha}^{\tilde{f}}$.
Minimization of the expectation value of $H_{\textrm{eff},0}$ leads to the following self-consistency 
conditions for $\lambda_{i}$ and $d_{i}$:
\begin{widetext}
\begin{eqnarray}
\lambda_{i} &=& \frac{1}{2} \biggl{(}  \frac{\partial g_{t,i\uparrow}}{\partial \bar{n}_{i\uparrow}^{\tilde{f}}} +  \frac{\partial g_{t,i\uparrow}}{\partial \bar{n}_{i\downarrow}^{\tilde{f}}} \biggl{)}
\sum_{j,\sigma \alpha}  [\mathbf{d}_{ji} \cdot \bm{\sigma}_{\sigma \alpha} \langle c_{j\sigma}^{\dagger} \tilde{f}_{i\alpha}\rangle + \textrm{c.c}] - \frac{3J_{H}}{4} \biggl{(} \frac{\partial g_{s,i}}{\partial \bar{n}_{i\uparrow}^{\tilde{f}}} + \frac{\partial g_{s,i}}{\partial \bar{n}_{i\downarrow}^{\tilde{f}}}\biggr{)}  \sum_{j} g_{s,j} \vert \tilde{\chi}_{ij} \vert^{2} \;, \label{eq:selfconsistency1} \\
-U_{f} &=& \sum_{j,\sigma\alpha} \biggl{[}  \mathbf{d}_{ji} \cdot \bm{\sigma}_{\sigma\alpha} \frac{\partial g_{t,i\alpha}}{\partial d_i}  \langle c_{j\sigma}^{\dagger}\tilde{f}_{i\alpha}\rangle + \textrm{c.c}\biggr{]}-\frac{3J_{H}}{2} \sum_{j} \frac{\partial g_{s,i}}{\partial d_i} g_{s,j} \vert \tilde{\chi}_{ij}\vert^{2} \;, \label{eq:selfconsistency2}
\end{eqnarray}
\end{widetext}
\noindent As it stands, the Lagranger multiplier $\lambda_i$ plays the role to counter-balance the bare-level effect on the $f$-electron occupation. Together with the spin liquid term, this ensures a finite Kondo renormalization. Equation~(\ref{EQ:Hamil_GA}) can be cast into the Anderson-Bogoliubov-de Gennes (Anderson-BdG) 
equations~\cite{JXZhu:2008,JXZhu:2016}:
\begin{equation}
\sum_{j} \left(
\begin{array}{cc}
h_{ij}^{c} \hat{I} &  \hat{\Delta}_{ij}   \\
\hat{\Delta}_{ij}^{\dagger}    & h_{ij}^{\tilde{f}} \hat{I}
\end{array} \right)
\left( \begin{array}{c}
\hat{u}_{j}^{n} \\
\hat{v}_{j}^{n}
\end{array} \right)  = E_{n}
\left( \begin{array}{c}
\hat{u}_{i}^{n} \\
\hat{v}_{i}^{n}
\end{array} \right) \;,
\label{EQ:BdG}
\end{equation}
subject to the constraints given by Eqs.~(\ref{eq:selfconsistency1}) and (\ref{eq:selfconsistency2}).
Here $\hat{I}$ is a $2\times 2$ identity matrix, $h_{ij}^{c} = -t_{ij}^{c} -\mu \delta_{ij}$,  
$\Delta_{ij,\sigma\alpha} = \mathbf{d}_{ij}\cdot \bm{\sigma}_{\sigma\alpha} g_{t,j\alpha}$, 
 and $h_{ij}^{\tilde{f}} =  ( \epsilon_{f}+\lambda_{i})  -\mu)
 \delta_{ij}  -3J_{H}g_{s,i}g_{s,j}\tilde{\chi}_{ij}/4$.   
After the self-consistency is achieved, one can then calculate the projected local density of states (LDOS) as defined by:
\begin{equation} 
(\rho_{i\sigma}^{c},\rho_{i\alpha}^{\tilde{f}}) = - \sum_{n} (\vert u_{i\sigma}^{n} \vert^2,  \vert v_{i,\alpha}^{n} \vert^{2} ) \frac{\partial f_{FD}(E-E_n)}{\partial E}\;,
\end{equation}
where the Fermi-Dirac distribution function $f_{FD}(E)=[\exp(E/k_{B}T)+1]^{-1}$.
Throughout this work, the quasiparticle
energy is measured with respect to the Fermi energy and the energy unit $t^c=1$
is chosen.

\subsection{Determination of the $Z_2$  topological indices} 
It has been proved~\cite{LFu:2007} that in an insulator with time-reversal and space-inversion symmetry, the topology is determined by parity properties at the eight high-symmetry points, $\mathbf{k}_{m}^{*}$, which satisfy $\mathbf{k}_{m}^{*}=-\mathbf{k}_{m}^{*}+\mathbf{G}$ with $\mathbf{G}$ being the reciprocal lattice vectors. In the three-dimensional simple cubic system, these vectors can be easily found to be $\mathbf{k}_{m}^{*} = (\pi/a) (n_1,n_2,n_3)$ with $n_{i}=0,1$.  The parity eigenvalue at these high-symmetry $\mathbf{k}_{m}^{*}$ points is given by:
\begin{equation}
\delta_{m} = \textrm{sgn}[\xi^{c}_{\mathbf{k}_{m}^{*}} - \xi^{f}_{\textrm{eff},\mathbf{k}_{m}^{*}}]\;,
\end{equation}
where $\xi^{c}_{\mathbf{k}}=-2t^{c}(\cos k_x a+ \cos k_y a+ \cos k_z a)-\mu$ is the single-particle energy dispersion for conduction electrons in the three-dimensional bulk while $\xi^{f}_{\textrm{eff},\mathbf{k}}= [-2t^{f}_{\textrm{eff},\perp}(\cos k_x a+ \cos k_y a) -2 t^{f}_{\textrm{eff},z} \cos k_z a] + \epsilon_{f} - \mu + \lambda$ with $t_{\textrm{eff},\perp(z)}=3J_{H}g_{s}^{2} \chi_{\perp(z)}/4$. The $Z_2$ topological indices can then be evaluated according to  $I_{\textrm{STKI}}=(-1)^{\nu_{0}}= \Pi_{m=1}^{8} \delta_{m}=\pm 1$ and $I_{\textrm{WTKI}}^{b}=(-1)^{\nu_{b}}=\Pi_{m} \delta_{m} \vert_{k_{m,b}^{*}=0}=\pm 1$ for $b=1,2,3$ corresponding to $x$, $y$, and $z$.~\cite{MDzero:2010} For a strong topological Kondo insulator $I_{\textrm{STKI}}=-1$ while for a weak topological Kondo insulator $I_{\textrm{WTKI}}^{b}=-1$.  By tuning $\epsilon_{f}$, we can get the two types of strong topological Kondo insulators with $Z_2$ indices $(\nu_{0};\nu_{1},\nu_{2},\nu_{3})=(-1;-1,-1,-1)$ and $(-1;0,0,0)$ and one weak topological Kondo insulator with $Z_2$ indices $(0;-1,-1,-1)$.

\subsection{Calculation of the local electron density of states around the single impurity}
Within the $T$-matrix approximation, the local electronic Green's function can be obtained as:
\begin{equation}
\hat{G}_{ij}(i\omega_n) = \hat{G}^{(0)}_{ij}(i\omega_n) + \hat{G}^{(0)}_{i,0}(i\omega_n)\hat{T}(i\omega_n) \hat{G}^{(0)}_{0,j}(i\omega_n)\;.
\end{equation} 
Here the bare Green's function for the pristine system is given by 
\begin{equation}
 \hat{G}^{(0)}_{ij}(i\omega_n) = \sum_{n} \frac{\hat{\phi}_{i,n}\otimes \hat{\phi}_{j,n}^{\dagger}}{i\omega_{n} -E_{n}}\;,
 \label{eq:G0_realspace}
 \end{equation}
 where  $\hat{\phi}_{i,n}^{\dagger}=(u_{i,\uparrow}^{n},u_{i,\downarrow}^{n},v_{i,\uparrow}^{n},v_{i,\downarrow}^{n})$ are the eigenstates of Eq.~(\ref{EQ:BdG}) corresponding to eigenvalues $E_n$.  The Matsubara frequency $\omega_{n}=\pi(2n+1)T$ with $n$ being an integer.  For the bulk system, there is a translational invariance along three spatial direction, and we can perform the Fourier transform such that 
 \begin{equation}
 \hat{G}^{(0)}_{ij}(i\omega_n) = \frac{1}{N_{L}} \sum_{\mathbf{k}} \hat{G}^{(0)}_{\mathbf{k}}(i\omega_n)e^{i\mathbf{k} \cdot (\mathbf{r}_{i} - \mathbf{r}_{j})}\;,
 \end{equation}
 with $N_L$ being the lattice size and 
 \begin{equation}
 \hat{G}^{(0),-1}_{\mathbf{k}}(i\omega_n)= \left( \begin{array}{cc} 
 (i\omega_{n}- \xi_{\mathbf{k}}^{c})\hat{I}      &   g_{t} \mathbf{d}_{\mathbf{k}}  \cdot \bm{\sigma} \\
 (g_t \mathbf{d}_{\mathbf{k}}  \cdot \bm{\sigma})^{\dagger}  &   (i\omega_{n} - \xi_{\textrm{eff},\mathbf{k}}^{f} )\hat{I}
 \end{array} \right)\;,
 \end{equation}
where $\hat{I}$ is a $2 \times 2$ identify matrix and  $\mathbf{d}_{\mathbf{k}} = V_{cf} (-i\sin k_xa, i\sin k_y a, 2i\sin k_z a)$. For the slab system, there is a translational invariance in the slab plane and we can use represent the bare Green's function in a mixed representation:
\begin{equation}
\hat{G}^{(0)}_{ij}(i\omega_n) = \frac{1}{N_{L,\perp}} \sum_{\mathbf{k}} \hat{G}^{(0)}_{i_z,j_z}(\mathbf{k}_{\perp},i\omega_{n}) e^{i\mathbf{k}_{\perp} \cdot (\mathbf{r}_{i,\perp} - \mathbf{r}_{j,\perp})}\;,
\end{equation}
where $N_{L,\perp}$ is the surface lattice size and 
\begin{equation}
 \hat{G}^{(0)}_{i_z,j_z}(\mathbf{k}_{\perp},i\omega_n) = \sum_{n} \frac{\hat{\phi}_{i_z,n}(\mathbf{k}_{\perp})\otimes \hat{\phi}_{j_z,n}^{\dagger}(\mathbf{k}_{\perp})}{i\omega_{n} -E_{n}(\mathbf{k}_{\perp})}\;,
 \label{eq:G0_mixedspace}
\end{equation}
with $\hat{\phi}_{i_z,n}(\mathbf{k}_{\perp})$ are the eigenstates of Eq.~(\ref{EQ:BdG}) but now written in the mixed representation with $\mathbf{k}_{\perp}$ a good quantum number.  Finally with the known bare Green's function, the $T$-matrix is given by 
\begin{equation}
\hat{T}(i\omega_{n})= [\hat{V}_{\textrm{imp}}^{-1} - \hat{G}^{(0)}_{oo}(i\omega_n)]^{-1}\;,
\end{equation}
with $\hat{G}^{(0)}_{oo}(i\omega_n)$ the local Green's function while $V_{\textrm{imp}}$ is the local impurity scattering matrix.  For the results reported in the paper, we take the intra-pseudospin channel scattering in the $f$-orbital with the strength of $U_{\textrm{imp}}$.

\subsection{Numerical details}
Throughout the work, the energy is measured with respect to the Fermi energy (i.e., chemical potential) and in units of the nearest-neighbor hopping integral $t^{c}=1$. The other parameter values are chosen as follows:  The temperature is fixed at $T=0.01$ to model the zero temperature limit, the strength of Hubbard repulsion  $U_{f}=15$ while $J_H=V_{cf}=1$. The electron filling factor is chosen to be $n_{\textrm{tot}}=1.95$. For our purpose to look into the local electronic structure around a single impurity on the surface, a slab geometry with a stack of 50 planes along the $z$-direction is considered. 

\medskip
\textbf{Supporting Information} \par 
Supporting Information is available from the Wiley Online Library or from the author.

\medskip
\textbf{Acknowledgements} \par 
This work was supported by the U.S. DOE NNSA under Contract No. 89233218CNA000001. It was supported by the U.S. DOE Office of Basic Energy Sciences Program  and the Los Alamos National Laboratory (LANL) LDRD Program. It was supported in part by the Center for Integrated Nanotechnologies, an Office of Science User Facility operated by the U.S. Department of Energy (DOE) Office of Science, in partnership with the LANL Institutional Computing Program for computational resources.  

\medskip
\textbf{Data Availability} \par 
The authors will make data available upon reasonable request.

\medskip
\textbf{Conflicts of Interest} \par 
The authors declare no conflicts of interest.


%
\bibliographystyle{MSP}

\begin{thebibliography}{10}
\providecommand{\url}[1]{\texttt{#1}}
\providecommand{\urlprefix}{URL }



\bibitem{MZHasan:2010}  M. Z.  Hasan  and  C. L. Kane,
Rev. Mod. Phys. {\bf 82}, 3045 (2010).

\bibitem{XLQi:2010} X.-L. Qi and S.-C.  Zhang, 
Phys. Today {\bf 63}, 33-38 (2010).  

\bibitem{XLQi:2011} X.-L. Qi and S.-C. Zhang, 
Rev. Mod. Phys. {\bf 83}, 1057 (2011).

\bibitem{YXia:2009} T. Xia, , D. Qian, D. Hsieh, L.  Wray,  A. Pal,  H.  Lin,  A. Bansil, D.  Grauer, Y. S.  Hor,   R. J. Cava,  and M. Z. Hasan, 
Nat. Phys. {\bf 5}, 398 (2009).

\bibitem{PRoushan:2009}  P.  Roushan, J.  Seo,  C. V.  Parker,  Y. S. Hor,  D. Hsieh,   D. Qian,  A. Richardella,  M. Z. Hasan,  R. J.  Cava, and  A. Yazdani,  
Nature (London) {\bf 460}, 1106 (2009).

\bibitem{TZhang:2009} T. Zhang,  P. Cheng, X. Chen,  J. F. Jia,  X. C. Ma,  K.  He, L. L.  Wang,  H. J. Zhang,  X. Dai, Z.  Fang,  X. C. Xie,  and Q. K.  Xue,  
Phys. Rev. Lett. {\bf 103}, 266803 (2009).

\bibitem{JSeo:2010} J. Seo,  P. Roushan,   H. Beidenkopf, Y. S.  Hor,  R. J.  Cava, and  A. Yazdani, 
Nature (London) {\bf 466}, 343 (2010).

\bibitem{ZAlpichshev:2010}  Z. Alpichshev,  J. G.  Analytis,   J. H. Chu,  I. R. Fisher,  Y. L.  Chen,  Z. X.  Shen, A.  Fang,  and  A. Kapitulnik,  
Phys. Rev. Lett. {\bf 104}, 016401 (2010).

\bibitem{BABernevig:2006} B. A. Bernevig,  T. L. Hughes,  and S. C. Zhang,  
Science {\bf 314}, 1757 (2006).

\bibitem{SRaghu:2008}  S. Raghu, X.-L.  Qi,   C. Honerkamp, and S.-C. Zhang,  
Phys. Rev. Lett. {\bf 100}, 156401 (2008).

\bibitem{KSun:2009}  K. Sun,   H. Yao,  E. Fradkin,  and S. A.  Kivelson,  
Phys. Rev. Lett. {\bf 103}, 046811 (2009).

\bibitem{RNandkishore:2010}  R. Nandkishore  and  L. Levitov,   
Phys. Rev. B {\bf 82}, 115124 (2010).

\bibitem{KSun:2012}  K. Sun,  W. V. Liu,  A. Hemmerich,  and S.  Das Sarma, 
Nat. Phys. {\bf 8}, 67 (2012).

\bibitem{HMGuo:2009} 
H. M. Guo  and   M. Franz,
Phys. Rev. Lett. {\bf 103}, 206805 (2009).

\bibitem{DAPesin:2010}  D. A. Pesin  and  L.  Balents,
Nat. Phys. {\bf 6}, 376 (2010).

\bibitem{XWan:2011}
X. Wan,   A. Turner,  A.  Vishwanath, and  S. Y.  Savrasov, 
Phys. Rev. B {\bf 83}, 205101 (2011).

\bibitem{BJYang:2010}
B.-J. Yang  and  Y. B. Kim, 
Phys. Rev. B {\bf 82}, 085111 (2010).

\bibitem{MDzero:2010}
M.  Dzero, K. Sun, V.  Galitski,  and P. Coleman, 
Phys. Rev. Lett. {\bf 104}, 106408 (2010).

\bibitem{XZhang:2012} X. Zhang,  H. Zhang, J. Wang, C. Felser,  and S.-C. Zhang,  
Science {\bf 335}, 1464 (2012).

\bibitem{BYan:2012}  B. Yan,   L. M\"{u}chler,   X.-L. Qi, S. C. Zhang,  and C.  Felser, 
Phys. Rev. B {\bf 85}, 165125 (2012).

\bibitem{MDzero:2012}  M. Dzero,  K. Sun,  P. Coleman,  and  V.  Galitski, 
Phys. Rev. B {\bf 85}, 045130 (2012).

\bibitem{MTTran:2012}  M.-T. Tran,  T. Takimoto,  and K.-S.   Kim, 
Phys. Rev. B {\bf 85}, 125128 (2012).

\bibitem{FLu:2013} F. Lu,  J.-Z. Zhao,  H. Weng, Z. Fang,  and  X.  Dai,  
Phys. Rev. Lett. {\bf 110}, 096401 (2013). 

\bibitem{XDeng:2013}  X. Deng,  K. Haule,  and  G.  Kotliar, 
Phys. Rev. Lett. {\bf 111}, 176404 (2013).

\bibitem{VAlexandrov:2013}
V. Alexandrov,  M. Dzero,  and  P. Coleman, 
Phys. Rev. Lett. {\bf 111}, 226403 (2013). 

\bibitem{XYFeng:2013} X.-Y. Feng, J. Dai,  C.-H. Chung,  and Q. Si,  
Phys. Rev. Lett. {\bf 111}, 016402 (2013).

\bibitem{TTakimoto:2011}  T. Takimoto, 
J. Phys. Soc. Jpn. {\bf 80}, 123710 (2011).

\bibitem{SWolgast:2013} S. Wolgast, \c{C}. Kurdak,  K. Sun,  J. W. Allen, D.-J. Kim,  and  Fisk,  Z.
Phys. Rev. B {\bf 88}, 180405(R) (2013).

\bibitem{ZJYue:2013} Z. J. Yue,  X. L. Wang, D. L. Wang, J. Y. Wang,  and S. X. Dou, 
J. Phys. Soc. Jpn. {\bf 84}, 044715 (2015).

\bibitem{GLi:2013} G. Li,  Z. Xiang,  F. Yu, T. Asaba, B.  Lawson, P. Cai,  C. Tinsman, A. Berkley, S.  Wolgast,  Y. S. Eo,  D.-J. Kim, C.  Kurdak,  J. W.  Allen,  K. Sun,  X. H. Chen, Y. Y.  Wang, Z.  Fisk,  L.  Li,  
Science {\bf 346}, 1208 (2014).

\bibitem{NXu:2013}  N. Xu,  X.  Shi, P. K.  Biswas,  C. E. Matt, R. S. Dhaka, Y.  Huang,  N. C. Plumb,  M.  Radovi\'{c},   J. H., Dil,  E. Pomjakushina, K.  Conder,   A. Amato,  Z. Salman, D. McK. Paul,   J. Mesot,  H.  Ding,  and  M.  Shi, 
Phys. Rev. B {\bf 88}, 121102(R) (2013).

\bibitem{MNeupane:2013} M. Neupane,  N. Alidoust,  S.-Y. Xu, T.  Kondo, D.-J. Kim,   C. Liu,   I.  Belopolski,  T.-R. Chang,  H.-T. Jeng, T.  Durakiewicz,  L. Balicas,  H. Lin, A. Bansil,  S. Shin,  Z. Fisk, and  M. Z. Hasan,  
Nat. Commun. {\bf 4}, 2991 (2013).

\bibitem{JJiang:2013}  J. Jiang,  S. Li,  T. Zhang, Z. Sun,  F. Chen,  Z. R. Ye, M.  Xu, Q. Q.  Ge,  S. Y. Tan, X. H.  Niu,  M. Xia, B. P. Xie,  Y. F. Li,  X. H. Chen,   H. H. Wen, and D. L. Feng, 
Nat. Commun. {\bf 4}, 3010 (2013).

\bibitem{MMYee:2013}  M. M. Yee, Y. He, A.  Soumyanarayanan, D.-J. Kim, Z. Fisk,  and J. E. Hoffman,  
arxiv.org:1308.1085.


\bibitem{ZHZhu:2013}
Z.-H. Zhu,  A.  Nicolaou,  G.  Levy,  N. P. Butch,  P.  Syers,   X. F.  Wang,
J. Paglione, G. A.  Sawatzky,  I. S.  Elfimov,  and A.  Damascelli, 
Phys. Rev. Lett. {\bf 111}, 216402 (2013).

\bibitem{EFrantzeskakis:2013}
E. Frantzeskakis,  N.  de Jong,  B. Zwartsenberg,  Y. K.  Huang,   Y.   Pan, X. Zhang,  J. X.   Zhang,  F. X. Zhang,    L. H. Bao,
O. Tegus,  A. Varykhalov,  A. de Visser,  and  M. S.  Golden,  
Phys. Rev. X {\bf 3}, 041024 (2013).

\bibitem{SRobler:2013}
S. R\"{o}$\beta$ler,  T.-H.  Jang, D. J.,  Kim,  L. H.  Tjeng,   Z.  Fisk,  F. Steglich, and S. Wirth, 
Proc. Natl. Acad. Sci. (USA ){\bf 111}, 4798 (2014). 

\bibitem{AVBalatsky:2006} A. V. Balatsky,  I. Vekhter,   J.-X. Zhu, 
Rev. Mod. Phys. {\bf 78}, 373 (2006).

\bibitem{JXZhu:2011} J.-X. Zhu,  R. Yu,   A. V.  Balatsky, and Q.   Si, 
Phys. Rev. Lett. {\bf 107}, 167002 (2011).

\bibitem{CLSong:2013} C.-L. Song, and J. E.  Hoffman, 
Curr. Opin. Solid State Mater. Sci.  {\bf 17}, 39 (2013).

\bibitem{MHHamidian:2011}
M. H. Hamidian,  A. R. Schmidt,  I. A. Firmo, M. P. Allan,  P. Bradley, J. D. Garrett,  T. J. Williams, G. M.,  Luke, Y. Dubi, A. V.  Balatsky,  and J. C. Davis, 
Proc. Natl. Acad. Sci. (USA) {\bf 108}, 18233 (2011).

\bibitem{MDzero:2012} M. O. Dzero, Eur. Phys. J. B {\bf 85}, 297 (2012).

\bibitem{TSenthil:2003} 
T. Senthil,  S. Sachdev,  and M.  Vojta, 
Phys. Rev. Lett. {\bf 90}, 216403 (2003).

\bibitem{TSenthil:2004}
T. Senthil,   M. Vojta,  and  S.  Sachdev, 
Phys. Rev. B {\bf 69}, 035111 (2004).

\bibitem{CPepin:2007}
C.  Pepin, 
Phys. Rev. Lett. {\bf 98}, 206401 (2007).

\bibitem{JXZhu:2008}
J.-X. Zhu,   I. Martin, and A. R.   Bishop, 
Phys. Rev. Lett. {\bf 100}, 236403 (2008).

\bibitem{PGegenwart:2008}
P. Gegenwart,  Q.  Si, and  F. Steglich,  
Nat. Phys. {\bf 4}, 186 (2008).

\bibitem{JXZhu:2012}
J.-X. Zhu,  J.-P. Julien, Y.  Dubi,  and  A. V. Balatsky, 
Phys. Rev. Lett. {\bf 108}, 186401 (2012).

\bibitem{LFu:2007} L. Fu,  and C. L.  Kane, 
Phys. Rev. B {\bf 76}, 045302 (2007).


\bibitem{MCGutzwiller:1963}  M. C. Gutzwiller,  
Phys. Rev. Lett. {\bf 10}, 159 (1963).

\bibitem{MCGutzwiller:1965}  M. C. Gutzwiller, 
Phys. Rev. {\bf 137},  A1726 (1965).

\bibitem{DVollhardt:1984}  D. Vollhardt, 
Rev. Mod. Phys. {\bf 56}, 99 (1984).

\bibitem{FCZhang:1988} F. C. Zhang,  
C. Gross,  T. M. Rice, and   H. Shiba, 
Supercond. Sci. Technol. {\bf 1}, 36 (1988).

\bibitem{CLi:2006}  C. Li,  S.  Zhou,  and  Z. Wang, 
Phys. Rev. B {\bf 73}, 060501(R) (2006).

\bibitem{JPJulien:2006}  J.-P.  Julien and  J. Bouchet, 
Prog. Theor. Chem. Phys. {\bf 15}, 509 (2006).

\bibitem{QHWang:2006}  Q.-H. Wang, 
Z. D. Wang,  Y.  Chen, and  F. C. Zhang, 
Phys. Rev. B {\bf 73}, 092507 (2006).

\bibitem{WHKo:2007}   W. H. Ko,  C. P. Pave,  and P. A.  Lee, 
Phys. Rev. B {\bf 76}, 245113 (2007).

\bibitem{NFukushima:2008}  N. Fukushima, 
Phys. Rev. B {\bf 78}, 115105 (2008).

\bibitem{JXZhu:2016} Jian-Xin Zhu, {\em Bogoliubov-de Gennes Method and Applictations} (Springer, Berlin, 2016).

\bibitem{OErten:2016} O. Erten, P. Ghaemi, and P. Coleman,  Phys. Rev. Lett. {\bf 116}, 046403 (2016).


\end{thebibliography}


\providecommand{\noopsort}[1]{}\providecommand{\singleletter}[1]{#1}%

\end{document}